\documentclass[twocolumn]{aastex631}

\usepackage{graphicx}
\usepackage{txfonts}
\usepackage{color}

\graphicspath{ {./Figures/} }

\usepackage{natbib}

\shorttitle{CASA on the fringe}

\def\rpicard{\textsf{\itshape rPICARD}}
\newcommand{\casatask}[1]{\texttt{\textbf{{#1}}}}
\newcommand{\aipstask}[1]{\texttt{\MakeUppercase{#1}}}
\newcommand{\casaparm}[1]{\texttt{\textit{#1}}}

\begin{document}

\title{CASA on the fringe -- Development of VLBI processing capabilities for CASA}

\author[0000-0001-5473-2950]{Ilse M. van Bemmel}
\affiliation{Joint Institute for VLBI ERIC (JIVE), Oude Hoogeveensedijk 4, 7991 PD Dwingeloo, The Netherlands}
\author[0000-0002-6156-5617]{Mark Kettenis}
\affiliation{Joint Institute for VLBI ERIC (JIVE), Oude Hoogeveensedijk 4, 7991 PD Dwingeloo, The Netherlands}
\author[0000-0003-3723-5404]{Des Small}
\affiliation{Joint Institute for VLBI ERIC (JIVE), Oude Hoogeveensedijk 4, 7991 PD Dwingeloo, The Netherlands}
\author[0000-0001-8685-6544]{Michael Janssen}
\affiliation{Max-Planck-Institut f\"ur Radioastronomie, Auf dem H\"ugel 69, D-53121 Bonn, Germany}
\author[0000-0002-3296-8134]{George A. Moellenbrock}
\affiliation{National Radio Astronomy Observatory, P.O. Box O, Socorro, NM 87801, USA}
\author[0000-0002-8704-7690]{Dirk Petry}
\affiliation{European Southern Observatory, Karl-Schwarzschild-Strasse 2, 85748 Garching, Germany}
\author[0000-0002-2542-7743]{Ciriaco Goddi} 
\affiliation{Dipartimento di Fisica, Universit\'a degli Studi di Cagliari, SP Monserrato-Sestu km 0.7, I-09042 Monserrato,  Italy}
\affiliation{INAF - Osservatorio Astronomico di Cagliari, via della Scienza 5, I-09047 Selargius (CA), Italy}
\author[0000-0002-3873-5497]{Justin D. Linford}
\affiliation{National Radio Astronomy Observatory, P.O. Box O, Socorro, NM 87801, USA}
\author[0000-0003-4146-9043]{Kazi L. J. Rygl}
\affiliation{INAF-Istituto di Radioastronomia \& Italian ALMA Regional Centre, Via P. Gobetti 101, I-40129 Bologna, Italy}
\author[0000-0003-0995-5201]{Elisabetta Liuzzo}
\affiliation{INAF-Istituto di Radioastronomia \& Italian ALMA Regional Centre, Via P. Gobetti 101, I-40129 Bologna, Italy}
\author[0000-0001-9814-2354]{Benito Marcote}
\affiliation{Joint Institute for VLBI ERIC (JIVE), Oude Hoogeveensedijk 4, 7991 PD Dwingeloo, The Netherlands}
\author[0000-0003-4116-4426]{Olga S. Bayandina}
\affiliation{Joint Institute for VLBI ERIC (JIVE), Oude Hoogeveensedijk 4, 7991 PD Dwingeloo, The Netherlands}
\affiliation{INAF -- Osservatorio Astrofisico di Arcetri, Largo E. Fermi 5, 50125 Firenze, Italy}
\author[0000-0002-3287-1535]{Neal Schweighart}
\affiliation{National Radio Astronomy Observatory, P.O. Box O, Socorro, NM 87801, USA}
\author[0000-0003-2884-9834]{Marjolein Verkouter}
\affiliation{Joint Institute for VLBI ERIC (JIVE), Oude Hoogeveensedijk 4, 7991 PD Dwingeloo, The Netherlands}
\author[0000-0002-5575-2774]{Aard Keimpema}
\affiliation{Joint Institute for VLBI ERIC (JIVE), Oude Hoogeveensedijk 4, 7991 PD Dwingeloo, The Netherlands}
\author[0000-0001-8525-4605]{Arpad Szomoru}
\affiliation{Joint Institute for VLBI ERIC (JIVE), Oude Hoogeveensedijk 4, 7991 PD Dwingeloo, The Netherlands}
\author[0000-0002-0230-5946]{Huib Jan van Langevelde}
\affiliation{Joint Institute for VLBI ERIC (JIVE), Oude Hoogeveensedijk 4, 7991 PD Dwingeloo, The Netherlands}
\affiliation{Leiden Observatory, Leiden University, Postbus 2300, 9513 RA Leiden, The Netherlands}
\affiliation{University of New Mexico, Department of Physics and Astronomy, Albuquerque, NM 87131, USA}

\begin{abstract}
New functionality to process Very Long Baseline Interferometry (VLBI) data has been implemented in the CASA package. This includes two new tasks to handle fringe fitting and VLBI-specific amplitude calibration steps. Existing tasks have been adjusted to handle VLBI visibility data and calibration meta-data properly. With these updates, it is now possible to process VLBI continuum and spectral line observations in CASA. This article describes the development and implementation, and presents an outline for the workflow when calibrating European VLBI Network or Very Long Baseline Array data in CASA. Though the CASA VLBI functionality has already been vetted extensively as part of the Event Horizon Telescope data processing, in this paper we compare results for the same dataset processed in CASA and AIPS. We find identical results for the two packages and conclude that CASA in some cases performs better, though it cannot match AIPS for single-core processing time. The new functionality in CASA allows for easy development of pipelines or Jupyter notebooks, and thus contributes to raising VLBI data processing to present day standards for accessibility, reproducibility, and reusability.
\end{abstract}


%
%

\section{Introduction}
\subsection{Background} 
For decades the Very Long Baseline Interferometry (VLBI) community primarily processed data using the National Radio Astronomy Observatory (NRAO) Astronomical Image Processing System (AIPS)  software package \citep{aips}. However, over the last decade the Common Astronomy Software Application  \citep[CASA;][]{casa}
has replaced AIPS for most {\em non}-VLBI applications in radio astronomy. Requiring users wishing to process VLBI observations to learn AIPS is a significant barrier to entry. In addition, with the changing hardware architecture and ever growing data volumes, AIPS is running into limitations that are increasingly harder to overcome in software. A particular challenge was the calibration of high-frequency, global VLBI observations carried out with the Event Horizon Telescope   \citep[EHT;][]{eht-m87-paper3}. 
To overcome this challenge the ERC-funded {\it BlackHoleCam} project \citep{BHC2017}, initiated the development of the CASA-VLBI functionality which led to the development of the first CASA-based  calibration pipeline for VLBI data \citep{rpicard,eht-m87-paper2}. 

This development was preceded by a detailed comparative study among the  main radio-interferometric data processing software packages currently in use. The comparison scored each package on its suitability to build a pipeline for VLBI observations by comparing reliability, flexibility, sustainability, user access and support. From this exercise, CASA and AIPS were found to be the best options, with CASA being the prime choice due to continuous and future software development and extensive support for users of large observatories such as ALMA and VLA (see Appendix~\ref{sec:software_comparison}). The development of the CASA VLBI functionality has in the meantime matured into a joint effort between the Joint Institute for VLBI ERIC (JIVE) in the Netherlands and NRAO in the US to improve the accessibility for (new) users of the European VLBI Network (EVN) and the Very Large Baseline Array (VLBA). In recent years also the option to run CASA in a Jupyterhub environment provides a many new benefits \citep{keimpema}.

Recently the CASA~6 series was released, the Python3-based distribution of the CASA package, available as a modular and a monolithic distribution. A detailed description of the CASA~6 software is presented in \cite{casarefpaper}, hereafter called the CASA reference paper. In this paper we describe the additional functionality required for the processing of VLBI data as it is implemented in CASA~6.4.1. It is therefore entirely complementary to the CASA reference paper.

\subsection{CASA calibration framework and VLBI}
The ongoing introduction of VLBI capabilities in CASA takes full advantage of the modularity of the general CASA calibration model. As described in more detail in the CASA reference paper, this calibration model is based on the Jones matrix formalism of the Hamaker-Bregman-Sault Measurement Equation \citep{hbs2}. Adding VLBI-specific terms (e.g., fringe fitting) mainly requires introducing calibration-type-specific specializations to the CASA calibration framework.  These specializations transparently inherit general interfaces and features, and need only implement their specific properties, including the details of their time- and frequency-dependence, the algebra for calculating their Jones matrix elements, specialized solvers for deriving solutions, and any other specifics.  Most VLBI-specific code is therefore well-isolated from existing calibration code, and adopts existing mechanisms for data storage (the MeasurementSet), data iteration for calibration solving and application, pre-calibration for solving, calibration solution storage, plotting, etc.  VLBI data processing may also make use of the extensive suite of existing calibration (complex gain, bandpass, instrumental polarization, etc.) and imaging tasks.  As a result of this development strategy, the VLBI-specific terms fit naturally into CASA's generalized (self-)calibration mechanism.  At the same time, VLBI-specific capabilities are fully-integrated by default and become seamlessly available to more general CASA processing contexts (e.g., ALMA, JVLA).    Additionally, the VLBI development exercise has had a favorable influence on some of the more general aspects of the CASA calibration infrastructure, including a mechanism for creating empty solution tables for any calibration type, solution cadence improvements, solution interpolation, definitions for ancillary calibration (e.g., gain curves), etc. 

\subsection{This paper}
This paper is organized as follows. 
In section 2 we describe the development of new tasks and updates of existing tasks. In section 3 we present the main use cases for CASA-VLBI, and a rudimentary comparison with AIPS is given in section 4. In section 5 we discuss future development plans, and we wrap up in section 6 with a brief summary.
Throughout the paper we will refer to CASA tasks in \casatask{bold}, task parameters in \casaparm{italic} and AIPS tasks in \aipstask{all caps}.

\section{Development}
This was not the first attempt to implement VLBI tools in CASA. Past efforts have ended prematurely mainly because they attempted to improve on the existing algorithm. The scope of this work therefore was to replicate the AIPS functionality as closely as possible, and optimize later. This makes the CASA implementation of certain VLBI-tasks slower compared to the AIPS implementation (see Section~\ref{sec:comparison}). However, paralellization through the CASA message-passing interface (MPI) implementation can overcome this problem.

We assessed the CASA calibration framework in order to identify the work needed to enable data processing as it is done in the EVN pipeline (see Section~\ref{subsec:instruments}). We started with the data products found in the EVN archive, and assessed all calibration steps. Imaging and further analysis was excluded from this work as these processes are highly dependent on the science goal. Polarisation calibration has also not been handled in detail, though CASA is based on the Measurement Equation and preserves all information needed for this step  \citep[for details see][]{marti}.

For EVN data all pipeline steps were found to have equivalent tasks in CASA except for the fringe fitting. For VLBA data it was found that CASA had no equivalent task for \aipstask{accor}\footnote{\aipstask{ACCOR} corrects amplitudes in cross-correlation spectra 
due to errors in sampler thresholds using measurements of auto-correlation spectra}, which is required for data processed in the Distributed FX (DiFX)  correlator \citep{difx}. For each of these steps, a new CASA task was developed: \casatask{fringefit} and \casatask{accor}.

Additional work was needed to ensure the proper handling of VLBI meta-data and smooth operations of the new tasks within the CASA calibration framework. This was done in close collaboration with the CASA development team to ensure that CASA would maintain pre-existing functionality.

\subsection{New task: \casatask{fringefit}}
\subsubsection{Solving}
The general strategy we employ is closely modeled on that of \cite{schwabcotton}, which is a description of the mathematical framework behind the AIPS \aipstask{fring} task. In a preliminary step, for each baseline to a given reference antenna, the visibilities are Fourier-transformed in frequency and time and the peak of this transform is used to identify candidate parameter values in the two-dimensional space of delay and delay rate. The height of the peak is used to calculate a signal-to-noise ratio, and antennas whose baseline to the reference antenna do not exceed a user-specified threshold are excluded from further steps and will be flagged when the resulting calibration table is applied to the data.

In a second (optional) step, these parameter estimates are refined using a nonlinear least-squares solver that uses all baselines to all stations whose signal-to-noise ratio exceeded the threshold used in the FFT stage. While Schwab and Cotton implemented a specialized least-squares solver for AIPS \citep{schwabcotton}, the CASA implementation uses an implementation from the GNU Scientific Library\footnote{\url{https://www.gnu.org/software/gsl/}}.

In keeping with CASA's Measurement Equation based approach to calibration the parameters solved are stored in a calibration table from which Jones matrices can be calculated on demand. The fringe fit results are stored in a new type of $G$-Jones table, customised for the specific parameters that need to be stored; which are the phase (a.k.a. secular phase), the delay (phase slope as function of frequency), and the delay-rate (phase slope as function of time) and a dispersive delay term to characterise ionospheric delays proportional to the square of the wavelength \citep{fringmemo}.

It is common practice in radio astronomy to perform a fringefit on a bright source for a short interval with each spectral window separately; this can characterise and allow correction of any instrumental delays between the bands. Since the corrections calculated for this effect are typically applied to the whole data set, it is convenient to be able to zero the delay-rate term in the parameters, which would otherwise be extrapolated in time and dominate the correction. Yet, the delay-rate should always be included in the solve step to maximise the signal to noise value of the data. The CASA task \casatask{fringefit} has been outfitted with a dedicated parameter \casaparm{zerorates} to do exactly this.

Additionally it is possible to control which of the delay, delay-rate and dispersive delay parameters are included in the fringefit solution. The solution for phase cannot be switched off. By default, delay and delay-rate are solved for and dispersive delay is not. The motivating use-case for this option was to allow delay to be omitted when fitting spectral line data, but the functionality is quite general.

As is the case throughout CASA's calibration framework, it is straightforward to use a source model (via the \texttt{MODEL\char`_DATA} column of the MeasurementSet), and source models can be imported from images produced via self-calibration in various formats.

Not every feature available in AIPS has been implemented at the time of writing; notable omissions are that it is not yet possible to ``stack" baselines, or combine correlations (LL and RR, for example) in a single fringefit stage for added sensitivity, but the \casatask{fringefit} task is under active development, and new features continue to be added.

\subsubsection{Applying and interpolating}
CASA's calibration framework includes generic methods to interpolate and apply solutions determined by the various calibration tasks to the data to be calibrated.  The solutions determined by the \casatask{fringefit} task differ from the other calibration tasks in the sense that the solutions include the time-derivative of the phase (phase rate) as well as the phase itself.  This time-derivative needs to be taken into account when interpolating (and extrapolating) fringefit solutions in time in order to resolve phase ambiguities.  This means the standard time interpolation implementation that just does a linear interpolation of the individual parameters of a solution is not sufficient.  Therefore CASA's interpolation framework was extended such that individual calibration classes that form the implementation of the various calibration tasks can override the time interpolation mechanism.  A ``rate aware'' interpolation mechanism was added to the \casatask{fringefit} implementation.  The interpolated phase  is formed by extrapolating the nearest solutions in time in either direction using the time derivative and taking the average of these extrapolated solutions weighted according to their distance in time.  This yields a smooth solution that correctly tracks the phase evolution in time as long as the phase rates of the different solutions are comparable. The delay is interpolated and applied independently.

\subsection{New task: \casatask{accor}} 
Not all VLBI correlators normalize the visibility amplitudes, which is a pre-requisite for proper amplitude calibration.  A new CASA \casatask{accor} task has been written to do this normalization which divides the visibilities by the average of the auto-correlations on a timescale that can be specified by the user.  But since the task uses the common CASA interfaces for data selection it not only provides the functionality of the AIPS \aipstask{accor} task, but also the functionality of the AIPS \aipstask{acscl} task\footnote{\aipstask{acscl} is similar to \aipstask{accor} but uses only the inner part of the spectral bandpass to avoid a bias from bandpass falloffs, and should be performed after applying any bandpass corrections.} which is needed to correct wide-band VLBA data.

\vspace{5mm}
\subsection{Upgrading other tasks}

The CASA \casatask{gencal} task already included code to use system temperature ($T_{rm sys}$) measurements to calibrate visibility amplitudes. This code made the assumption that such measurements are made at the same cadence for all antennas in the array.  This is almost never the case for VLBI arrays so the code was changed to allow for measurements with a more irregular pattern by setting the parameter \casaparm{uniform} set to True (the default value).  In addition support was added for storing gain curves (which describe the elevation-dependency of the antenna gain) in the Measurement Set and using these in the \casatask{gencal} task to calibrate visibility amplitudes.

Many VLBI arrays use the standardized FITS Interferometry Data Interchange Convention (FITS-IDI; \citealt{aips102, aips114}).  In preparation for future VLBI work, a CASA task \casatask{importfitsidi} had already been developed by the CASA team for CASA~3/4 in 2010-12 as part of the European RadioNet-funded ALBiUS project. This task converts FITS-IDI data into a MeasurementSet. 
The initial implementation did not include a number of optional (according to the format definition) metadata tables. Further upgrades were made to the task in later CASA versions, and the current version will correctly import system temperature measurements, gain curves and weather data from FITS-IDI (see also section~3). Functionality to import the correlator model and pulse-cal measurements is still missing, but these measurements are not essential for the majority of VLBI observations.  Another important addition is that the \casatask{importfitsidi} task will now apply the so-called digital corrections required for DiFX correlated data in the same way as the AIPS \aipstask{fitld} task. These corrections are necessary to correct the effects of the coarse (2- or 4-level) signal quantization used by most VLBI data acquisition systems (generalisation of the Van-Vleck correction as described in Appendix 8.3 of \citealt{tms}). These corrections are non-linear in the correlation amplitude and require information not available to CASA’s generalized calibration framework, which is why the status quo of applying these in the “filler” has been retained. The current implementation applies this correction for 2- and 4-level quantization, and combinations thereof.


\subsection{Supporting scripts} \label{subsec:supportscripts} 
The traditional way to calibrate visibility amplitudes for connected-element radio interferometers is to observe a point-like source with known brightness and adjust the antenna gains to match the corresponding expected model visibilities.  In CASA this is done using the \casatask{setjy} and \casatask{gaincal} tasks.  This strategy does not work very well for VLBI since most bright calibrators are variable or have extended structure at the scales probed by VLBI.  Instead visibility amplitudes are calibrated based on system temperature ($T_{sys}$) measurements at the individual antennas and the antenna gain \citep{moran95}. Antenna gains are receiver-dependent and usually dependent on elevation as well.  Ideally both $T_{sys}$ and gain curves are distributed within the FITS-IDI data files.  Unfortunately many VLBI arrays distribute this meta-data separately, usually in AIPS ANTAB format.  To make this meta-data available to CASA a set of scripts were developed that add $T_{sys}$ measurements and gain curve information to FITS-IDI files that lack them.  These scripts are distributed separately from CASA.\footnote{\url{https://github.com/jive-vlbi/casa-vlbi}}

\section{VLBI applications} 
The CASA-VLBI functionality allows for calibration of the majority of VLBI observations. In this section we describe a few science cases for the EVN and VLBA. These instruments were chosen because of the differences in the processing steps caused by the differences in the correlators they use. The processing for continuum and spectral lines is discussed. Time domain and geodesy require a fundamentally different functionality (see also section \ref{sec:future}).

\subsection{Instruments} \label{subsec:instruments}
The majority of observations with the EVN is correlated at the Joint Institute for VLBI ERIC (JIVE; the Netherlands) and the pre-processed data are made available through the EVN Data Archive in FITS-IDI format\footnote{The EVN Data Archive is located at \url{http://archive.jive.nl/scripts/portal.php}}. The a-priori calibration information (gain and system temperature) as well as observational flags for all antennas can be applied by using the supporting scripts detailed in \S~\ref{subsec:supportscripts}. The visibility data can be imported in CASA with the task \casatask{importfitsidi}. The full calibration and imaging can then be performed within CASA with consistent results with respect to the calibration done in AIPS. JIVE provides a Jupyter Notebook for continuum calibration of any EVN observation, and a full guide to calibrating EVN data within CASA is also available at the EVN website \footnote{\url{https://www.evlbi.org/evn-data-reduction-guide}.}.

Observations with the Very Long Baseline Array (VLBA) can also be calibrated with CASA. VLBA observations are correlated with the DiFX correlator \citep{difx} at the NRAO Pete V. Domenici Science Operations Center (Socorro, NM, USA) and the correlated data are available through the NRAO Archive Access Tool\footnote{\url{https://data.nrao.edu/}}.  As of CASA~5.8/6.2, the CASA task \casatask{importfitsidi} is able to import the gain curve and system temperature tables from the FITS-IDI file.  The CASA \casatask{accor} task is capable of reproducing the behavior of the AIPS task \aipstask{acscl}, allowing for better amplitude calibration of wideband VLBA data.  A guide to calibrating VLBA data with CASA is available \citep{linford21}, and a CASA Guide tutorial is in development. Astronomers at the US Naval Observatory (USNO) have developed a pipeline for calibrating and imaging VLBA observations with CASA and have demonstrated that the results compare favorably with calibration done in AIPS and imaging done in Difmap \citep{hunt21}.

The flexibility of the CASA framework and its convenient interface to the MeasurementSet data makes it easy for users to build high-performance data processing pipelines with a Python front end.
In \cite{rpicard} \rpicard{} is presented, the first generic VLBI calibration and imaging pipeline that has been built fully on top of CASA and makes use of the new VLBI functionalities presented in this paper.
With \rpicard{}, the first VLBI data set fully calibrated and imaged with CASA has been published \citep{rpicard}. The M\,87 jet has been observed with the VLBA at 43\,GHz and imaged with the \casatask{tclean} task and CASA self-calibration methods implemented in \rpicard{}.
The pipeline is also used as a calibration pathway for the Event Horizon Telescope \citep[e.g,][]{eht-m87-paper3, cena-eht, eht-sgra-paper2}. 
With a significantly shorter track record, and continuous development, CASA performed as well as HOPS, the standard package for high-frequency VLBI calibration for the past decade \citep{hops, ehthops}.
More broadly, the calibration pipeline has been successfully applied to the EHT, GMVA, VLBA, EVN, and synthetic VLBI data from {\textsf{\itshape MeqSilhouette}} \citep{meqsil,symba, meqsil2}. Input files for these use-cases and a detailed documentation are available in the online repository of the \rpicard{} pipeline\footnote{\url{https://bitbucket.org/M_Janssen/picard}}.

\subsection{Main use cases}
The continuum data calibration in CASA follows the standard steps, which are described in the EVN User Manual and the VLBA User Manual (see also section~3.1). A notable difference with AIPS is that CASA writes the calibration solutions to an external calibration table, which is a separate directory on disk, and not a table associated with the {\it uv}-data. It is up to the user to specify which calibration tables to use when on-the-fly calibration is performed in tasks like \casatask{fringefit}. There is no cumulative calibration table in CASA, after all calibration steps are complete, the task \casatask{applycal} is used to apply all the calibration tables to the data.

Spectral line data in principle require the same calibration steps as continuum data, although they introduce a few additional complications. Spectral line observations require high spectral resolution, and the EVN correlator provides either one or two correlation passes. In the latter case, a separate low spectral-resolution pass is provided for calibrator sources. In the case of one correlator pass, the calibrator data needs to be manually averaged down to optimise signal to noise and processing time.
Spectral line calibration requires an excellent bandpass correction, especially in the case of HI-absorption line studies, which is possible with the very flexible CASA \casatask{bandpass} task. The \casatask{mstransform} can be used to re-reference the data in frequency domain. Self-calibration of a target maser source as well as inverse phase referencing is usually performed using only a few narrow spectral channels. This functionality is also available in CASA \casatask{fringefit}.

Flagging can be done using the automated \casaparm{rflag} option in the CASA \casatask{flagdata} task, by hand, or with any additional package. When flagging by hand, it is recommended to maintain a CASA flagmanager file to reproduce the flags at a later stage. The calibration tables can also be flagged, though it is better practice to flag the data that gives rise to poor calibration solutions, as this is generally bad data anyway. Particular attention has to to be paid to the target flagging steps of spectral line observations, as narrow spectral lines (especially in masers) can be confused with man-made radio frequency interference (RFI) by automatic flagging algorithms. In this case manual flagging is recommended.
After applying all the calibration tables to the data, the user can continue with imaging using CASA \casatask{tclean} or another package of their choice. We recommend to take care when changing the data format, and verify that all calibration and flags are properly applied in the new format before proceeding.

During the entire process of calibration, frequent quality control is necessary. The CASA task \casatask{plotms} can plot both visibility data and calibration solutions, also for the VLBI specific tables. As the CASA task \casatask{plotcal} is no longer included in CASA~6, the EVN Jupyter notebook environment uses a custom-made replacement called \casatask{plotcalng}\footnote{\url{https://github.com/aardk/evn-tools}} as the Matplolib-based plots produced by \casatask{plotcalng} are better suited for inclusion in a Jupyter notebook. Note that \casatask{plotcalng} is not part of the CASA package.

\begin{figure*}[t]
\vspace*{-15mm}
\hspace{-5mm}
\includegraphics[scale=0.5]{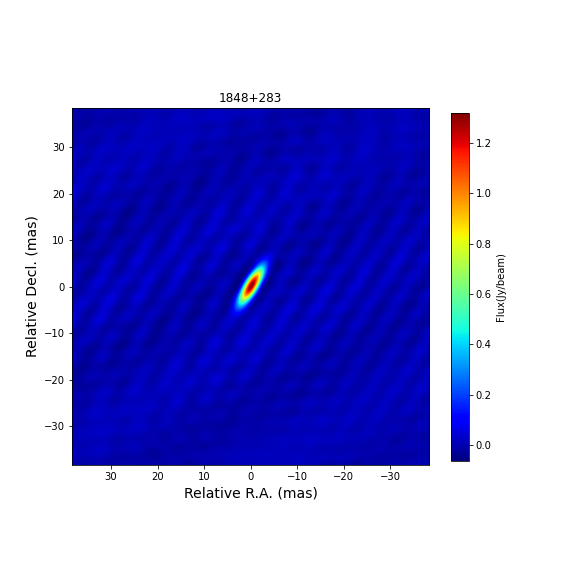}
\hspace{-10mm}
\includegraphics[scale=0.5]{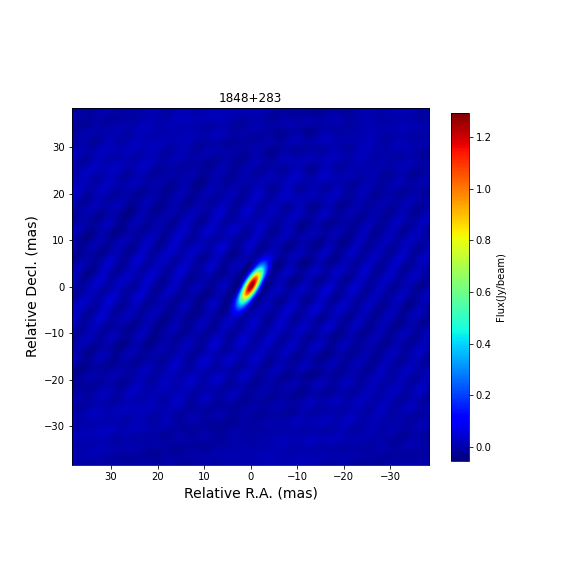}
\vspace*{-15mm}
\caption{Total intensity images of calibrator 1848+283. Data calibrated in AIPS (left) and CASA (right). Both calibrated data sets were imaged with \casatask{tclean} using the same parameters.}
\label{fig:1848+283_aips_image}
\end{figure*}

\begin{figure}[t]
\vspace{-12mm}
\hspace{-5mm}
\includegraphics[scale=0.5]{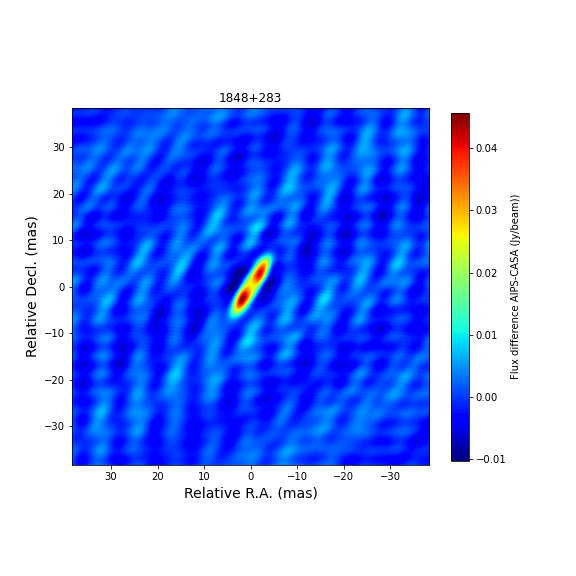}
\vspace{-15mm}
\caption{Difference image of the total intensity for 1848+382 (AIPS $-$ CASA). Note the narrow range of the colour scale.}
\label{fig:1848+283_diff_image}
\centering
\end{figure}

\section{Comparison}
\label{sec:comparison}
The EHT data calibration pipelines \citep{eht-m87-paper3} have demonstrated extensively that CASA performs as well as HOPS \citep{hops}  and AIPS \citep{aips}. The EHT data are difficult to calibrate compared to data from better understood instruments such as EVN and VLBA. A comparison with VLBA observations is done in \cite{hunt21}, finding no differences in the results. We include here a basic comparison for EVN observations, which observes at lower frequencies than EHT, and is a more in-homogeneous array than the VLBA.

A comparison between AIPS and CASA calibration was performed using EVN observations from the Network Monitoring Experiment with project code N14C3. This observation has been extensively used in testing and verification, and all outcomes and issues are known and understood. The observation includes a phase-referencing setup which is a standard continuum observation technique for VLBI. The data include a fringe finder source, and two pairs of a target and phase reference source. All sources are bright, which makes this dataset ideal for testing the frequently used phase-referencing mode of the EVN.

The data were taken in 2014, in C band. The participating telescopes were Effelsberg, a single dish from the Westerbork array, Jodrell Bank Lovell telescope, Onsala, Toru\'{n}, Svetloe, Zelenchukskaya, Badary, Sheshan, Hartebeesthoek and Yebes. The data were correlated in eight spectral windows of 32 channels each, and include all four polarization products. 
We used AIPS 31DEC22 \citep{aips}, ParselTongue3 \citep{pt} and CASA 6.4.1 \citep{casa, casarefpaper}. Two scripts were developed, one for AIPS and one for CASA and are publicly available at the JIVE code repository\footnote{\url{https://code.jive.eu/bemmel/Comparison_AIPS_CASA}}. The scripts use Python 3.8. 
The CASA script is based on the EVN Jupyter notebook for continuum data processing\footnote{\url{https://code.jive.eu/bemmel/EVN_CASA_pipeline}}, and the EVN Users Manual (see section~3.1). The AIPS script also follows the online EVN Users Manual, but in places where the order of steps or the parameters deviate from CASA, we give priority to the CASA version and change the AIPS settings to reflect the CASA procedure as closely as possible.

Both scripts handle the visibility calibration, using the standard steps (in this order): system temperature and gain corrections, flagging of data based on the telescope logs and known bad stations, fringe fitting for the delay (also referred to as instrumental delay), fringe fitting for the delay-rate (also referred to as multi-band delay), and finally a complex bandpass correction. 

Though we note that AIPS is generally faster than CASA, the processing speed of the two packages is difficult to compare due to differences in architecture, calibration model, and data format. These make CASA more user friendly, and more flexible for larger and complex datasets, at the expense of processing speed and memory use. For example, converting the data from FITS-IDI to MeasurementSet format is very time consuming, but typically this is only done once in the calibration procedure. To give a rough idea of the difference, we compared the processing speed of the fringe fitting step, where the implementation in the two packages is fairly similar, and for the entire calibration process. On a personal laptop CASA on a single core is a factor $\sim15-20$ slower in fringe fitting. For the full calibration procedure a similar difference is measured. When using the MPI option, this difference reduced to a factor $\sim2$, see Appendix~F in \cite{rpicard}.

\subsection{Images}
The final calibrated data are split into separate, calibrated datasets for each calibrator and target. From AIPS the calibrated data are exported to UV-FITS and then imported into CASA using \casatask{importuvfits}. Since we do not wish to compare the imaging tools, we choose to image all the datasets with CASA \casatask{tclean}. The imaging is done with a mild clean of 100 iterations, a cell size of 0.3\,mas, image size of 512 x 512 pixels and a clean box of 20 pixels in RA and DEC centered on the central pixel. The other parameters are set to the task defaults. This ensures that any differences in the resulting images are solely due to the calibration process. Since producing science quality images of VLBI observations is done in many different ways depending on the science goal, experience, and taste of the user, we perform no self-calibration or further imaging. All sources were imaged. We found the signal to noise ratio was best for the calibrator 1848+283, and therefore use this source for the further comparison. The reported trends are seen in all sources, but most obvious in the bright and compact calibrators.

In Fig.~\ref{fig:1848+283_aips_image} the resulting images are shown of the calibrator source 1848+283. For the AIPS calibrated data the peak brightness is $1.32 \pm 0.05\,\mathrm{Jy\ beam^{-1}}$, for the CASA calibrated data the peak flux is $1.30 \pm 0.05\,\mathrm{Jy\ beam^{-1}}$. The RMS noise in the AIPS image is 0.054\,Jy, versus 0.051\,Jy in the CASA image.

\begin{figure}[t]
\vspace{-4mm}
\includegraphics[scale=0.45]{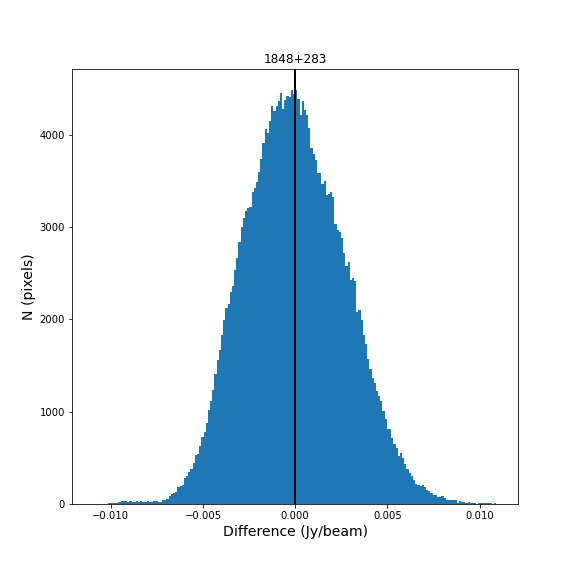}
\vspace{-6mm}
\caption{Histogram of the pixel flux in the AIPS $-$ CASA difference image.}
\label{fig:1848+283_diff_hist}
\centering
\end{figure}

Though the images look identical at first glance, a subtraction of the two reveals an extended residual (see Fig.~\ref{fig:1848+283_diff_image}). The same pattern is visible in the background at a much lower level. The residual flux is well below the RMS noise level of the original images. Though the majority of pixels shows a positive flux in this difference, implying that the AIPS flux calibration overall results in a higher flux, the distribution of the pixel values in the difference image is Gaussian and does not show a distinctive skew or offset (see Fig.~\ref{fig:1848+283_diff_hist}). This implies that the residuals originate from visibilities with similar noise realisation, but with a small (few percent) difference in the scaling.

\begin{figure*}[t]
\hspace{-5mm}
\includegraphics[scale=0.33]{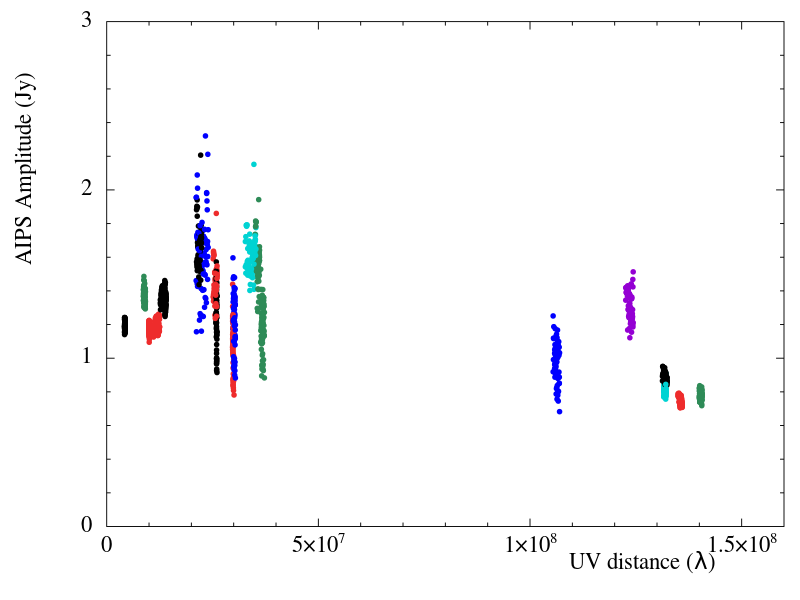}
\includegraphics[scale=0.33]{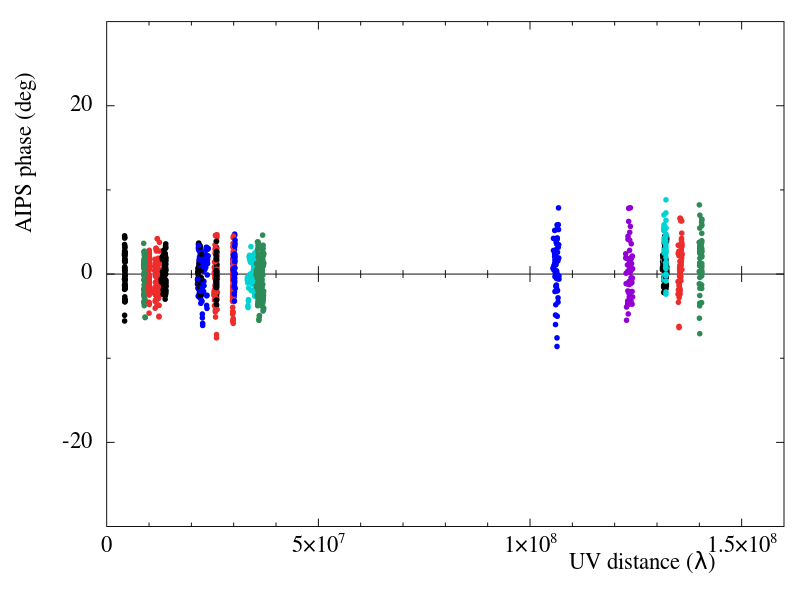}

\hspace{-5mm}
\includegraphics[scale=0.33]{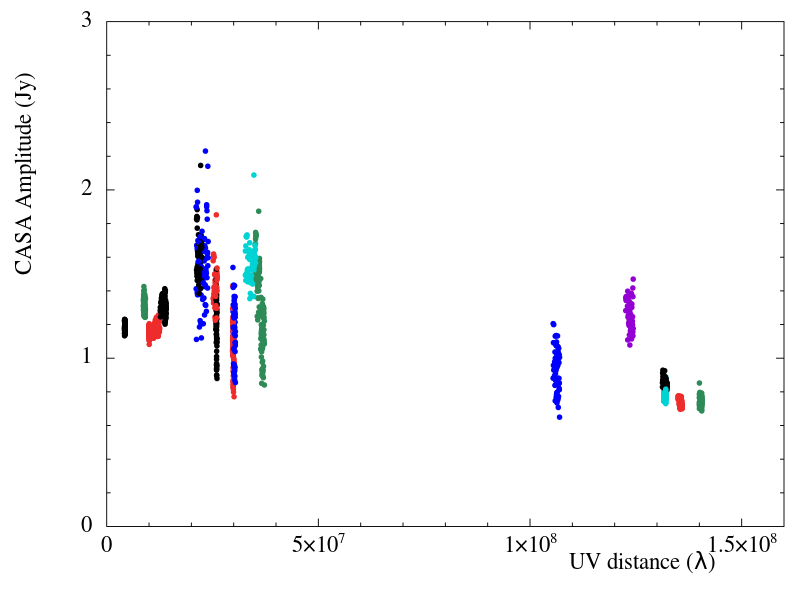}
\includegraphics[scale=0.33]{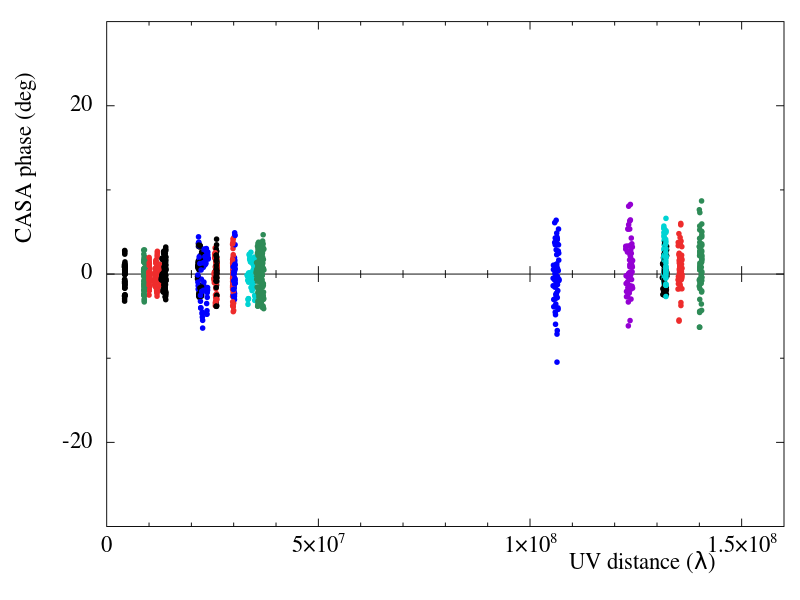}
\caption{Plots of calibrated amplitudes (left) and phases (right) as function of uv-distance for the calibrator 1848+283. The data include LL and RR polarization, and spectral windows 5 and 6. Top row is AIPS calibration, bottom row is CASA calibration. The colours indicate the first station in the baseline. Black: Effelsberg; red: Westerbork; dark green: Onsala; blue: Noto; cyan: Torun; purple: Hartebeesthoek.}
\label{fig:1848+283_aips_casa_uv}
\centering
\vspace{5mm}
\end{figure*}

\begin{figure*}[t]
\hspace{-5mm}
\includegraphics[scale=0.33]{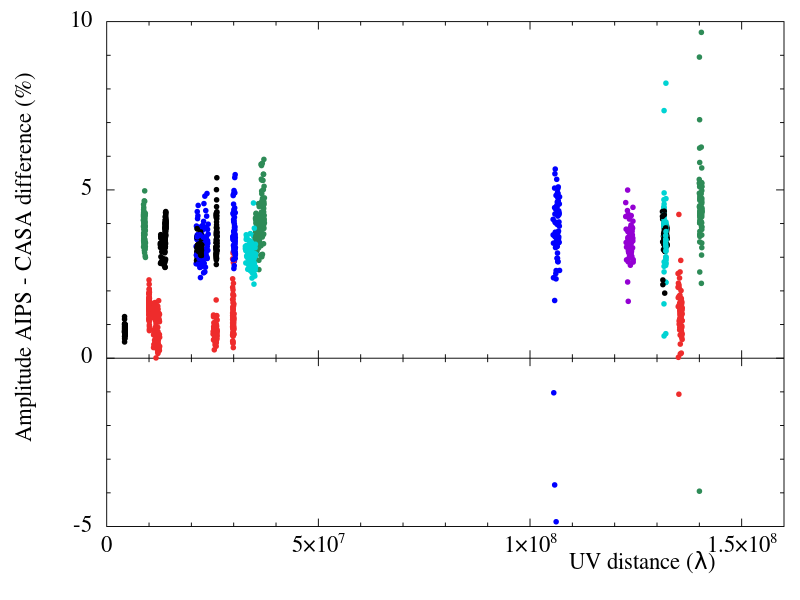}
\includegraphics[scale=0.33]{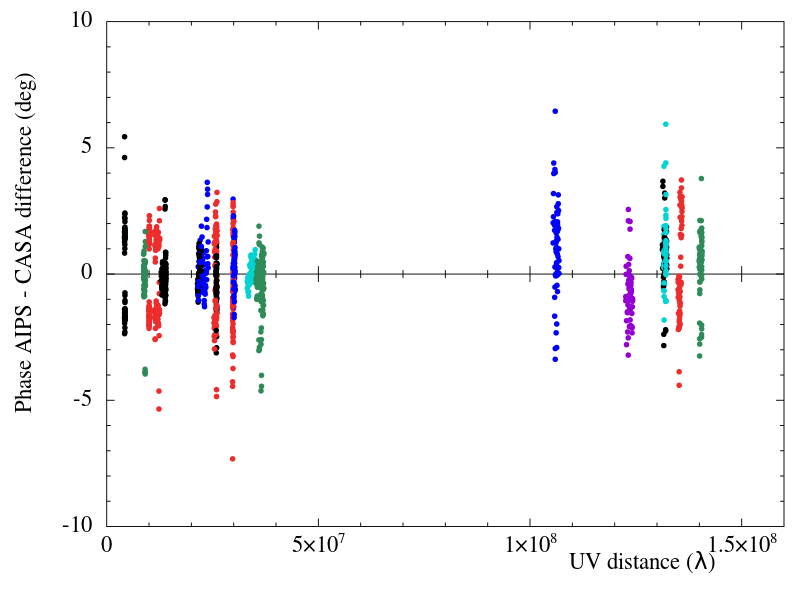}
\caption{Difference of the calibrated amplitudes (left) and phases (right) for the calibratoer 1848+283, using CASA $-$ AIPS as a function of uv-distance. The data include LL and RR polarization, and spectral windows 5 and 6. The colours indicate the first station in the baseline. Black: Effelsberg; red: Westerbork; dark-green: Onsala; blue: Noto; cyan: Torun; purple: Hartebeesthoek.}
\label{fig:1848+283_diff_uv}
\centering
\vspace{5mm}
\end{figure*}

\subsection{Amplitudes and phases}

To compare the calibration quality without imaging the data, plots are generated of the calibrated amplitudes and phases as a function of $uv$-distance. Figure~\ref{fig:1848+283_aips_casa_uv} shows the AIPS  and CASA calibration results. Individual data points are plotted for each scan on 1848+283, both LL and RR polarizations, and for spectral windows number five and six (in CASA zero-based counting). For clarity not all spectral windows are included. 

As for the total intensity images the differences between the AIPS and CASA calibration are not visible by eye, and a difference image was produced (see Fig.~\ref{fig:1848+283_diff_uv}). This demonstrates the same flux offset as seen in the total intensity images: the AIPS amplitudes are consistently higher than the CASA amplitudes by a few percent. This is somewhat larger than the differences found by \cite{hunt21}, and we ascribe this to manual editing of the gain information in the archival data. Both packages should take their gain and system temperature information from the archival ANTAB file, however, for AIPS this resulted in errors in the gain calibration. This was overcome by using the calibration information in the AIPS CL2 table provided in the EVN archive. The generation of that particular CL2 calibration table cannot be reproduced with the current version of AIPS. 
The phases differences are all clustered around $0^\circ$ phase, with at most a few degrees offset. The absolute amplitude and phase values are well within the empirically established accuracy limits for self calibration of 10\% amplitude errors and $10^\circ$ phase errors. 
The baselines with Westerbork show a split in the phase differences, which has been traced back to the instrumental delay calibration in AIPS. The odd and even spectral windows have respectively a positive and negative phase offset after the instrumental delay has been applied. These correspond to upper- and lower-sideband data, and indicate a small phase difference between those two, which is not properly corrected in AIPS. The offset is no more than $2^\circ$ in either direction, a value much smaller than typical residual source errors. The fact that this offset is spotted in the phase differences indicates that the quality of the calibration in both packages is comparable and accurate to within less than a degree. 

\vspace{1cm}
\section{Future plans}
\label{sec:future}
While the functionality implemented in CASA 6.4 should be sufficient to achieve the scientific goals of many types of VLBI observation, some more advanced functionality is still missing.  Work continues to implement this. Here we review several upcoming changes and improvements. Note that this list is not complete, and as instruments advance, the software development will follow.
\begin{itemize}
\item {\bf Pulse cal tones.}  This may help taking out instrumental delays by making use of calibration signals inserted at the receivers and measured at the VLBI correlator.
\item {\bf Fringe fit improvements.}  Several improvements are planned such as baseline stacking, stacking of polarisation products, and multi-band fringe fitting of irregularly spaced spectral windows.
\item {\bf Polarisation calibration.}  Full-stokes imaging requires calibration of the polarisation leakage (the so-called D-terms).  For polarisation observations with VLBI instruments the calibration algorithm needs to be able to handle resolved polarisation calibrator sources.  This is something that CASA currently does not provide.  A possible candidate software package is PolSolve \citep{marti} since it fully exploits the Measurement Equation.
\item {\bf Ionospheric corrections.}  While CASA implements ionospheric corrections in the \casatask{gencal} task, preliminary evaluation of this task suggests that improvements can be made for VLBI.
\item {\bf Delay model accountability and EOP corrections.}  For astrometric VLBI observations it is desirable to be able to make corrections to the delay model used by the correlator.  The most important of these are corrections to the Earth Orientation Parameters (EOPs) since the most accurate measurements of these parameters are only available some time after observations.
\end{itemize}

Our focus has been primarily on processing VLBI radio astronomy observations.  But there are opportunities for using CASA for calibration of geodetic observations as well.  The \casatask{fringefit} task is well-positioned to handle the increased bandwidth available in the VGOS system \citep{vgos}.  And plans to include resolved sources in geodetic observations will require the capability to fringe fit based on a source model, as well as the imaging capabilities to determine a source model in the first place.  CASA already provides that functionality, unlike the HOPS package that is currently used to process most geodetic observations.  How to reconcile the antenna-based solutions from a global fringe fitter with the baseline-based solutions required by geodetic analysis software is an open question though. Note that the CASA fringe fitter has a mode to skip the least-squares globalization step and provide just the baseline-based solutions in a calibration table, but it does not currently convert this into the information needed for geodetic experiments.

While CASA is undeniably based on more modern technologies than AIPS, its data access layers provided by \textsf{\itshape{casacore}} are limiting its scalability.  This already is an issue for larger observations with current instruments but will be a problem for future instruments such as ngVLA.  At NRAO, the CASA Next Generation Infrastructure (CNGI) project studied new technologies for replacing \textsf{\itshape{casacore}} and the existing MPI-based paralelization framework.  The conclusion from this study is that Python-based Xarray\footnote{\url{https://docs.xarray.dev/en/stable/}} and Dask\footnote{\url{https://docs.dask.org/en/stable/}} technologies are good choices to develop a Next Generation CASA (ngCASA).  The current plan is to introduce these technologies step by step into CASA.  While scalability is less of an issue for typical VLBI observations, more demanding VLBI observations (such as wide-field observations) can benefit from these new technologies as well.  Therefore we expect to adopt these technologies in future developments of VLBI-specific CASA tasks and ultimately reimplement already existing tasks (such as \casatask{fringefit}) on top of these new technologies \citep[see][]{casarefpaper}. 

\section{Summary}
The implementation of VLBI functionality in CASA is an important step towards sustainable software for VLBI data processing. It also opens the route towards a Findable, Accessible, Interoperable and Reusable (FAIR) software and data policy through Jupyter notebooks\footnote{\url{https://code.jive.eu/bemmel/EVN_CASA_pipeline}} that can be linked to the science data from a specific experiment, improving scientific reproducibility as well. This paper presents the current status of the VLBI functionality of the CASA package, based on CASA~6.4.1. New tasks have been added to handle the fringe fitting and specific amplitude calibration steps for VLBI use cases. Existing tasks have been updated to handle the calibration meta-data properly, and underlying processing has been adjusted to work in the extreme limits of VLBI observations. 

The CASA package is now capable of handling the majority of VLBI science cases, and has already undergone intense verification as part of the EHT data processing. The Python3 base allows for easy development of automated pipelines, and there is a Jupyter notebook kernel including CASA, which is extremely suitable for training purposes and less experienced users. We have presented an outline of the work flow for continuum and spectral line observations in CASA. Detailed recipes are available for the EVN and VLBA on their websites, and other telescopes are starting to provide similar resources.

A detailed comparison with the AIPS package has demonstrated that CASA performs equally well, though slower on a single CPU core. This can be overcome by using the MPI infrastructure, which enables multi-threading and can speed up the processing by orders of magnitude, depending on the underlying hardware.

As the technology of VLBI continues to evolve, the software is also under constant development. We have listed several improvements that are planned for the coming year. Longer term plans are developed jointly between JIVE and NRAO, and involve feedback from users and other stakeholders. 

The CASA VLBI functionality is not only suitable for radio astronomy applications, but can potentially also serve geodetic experiments. In the more distant future, the need for handling large data will require a significant overhaul of the underlying CASA infrastructure, and plans for this are already under development. We expect that the functionality presented in this paper will remain available. \vspace{1cm}


\section*{Acknowledgements}
We would like to thank Sophia Vaughan for her contribution in testing specific functionality of the fringefit task for spectral line applications during her Summer Studentship at JIVE in 2017. We thank the anonymous referee for a prompt and helpful assessment of the paper.
The National Radio Astronomy Observatory is a facility of the National Science Foundation operated under cooperative agreement by Associated Universities, Inc.
The European VLBI Network is a joint facility of independent European, African, Asian, and North American radio astronomy institutes. Scientific results from data presented in this publication are derived from the following EVN project code(s): N14C3.
This work is supported by the ERC Synergy Grant “BlackHoleCam: Imaging the Event Horizon of Black Holes” (Grant 610058).
This work is also supported by ESCAPE - The European Science Cluster of Astronomy \& Partcle Physics ESFRI Research Infrastructures, which has received funding from the European Union’s Horizon 2020 research and innovaton programme under the Grant Agreement 824064.
B.M. acknowledges support from the Spanish Ministerio de Econom\'ia y Competitividad (MINECO) under grants AYA2016-76012-C3-1-P and MDM-2014-0369 of ICCUB (Unidad de Excelencia ``Mar\'ia de Maeztu'').

\vspace{20mm}
\appendix
\section{Software comparison}
\label{sec:software_comparison}
A detailed comparison of several software packages was done in late 2014 as part of the requirements analysis for a pipeline that could process high-frequency VLBI observations from the GMVA and EHT. The software packages assessed were AIPS, Miriad\footnote{\url{https://www.atnf.csiro.au/computing/software/miriad/}} \citep{miriad}, HOPS\footnote{\url{https://www.haystack.mit.edu/haystack-observatory-postprocessing-system-hops/}}, \citep{hops, ehthops}, CASA \citep{casa, casarefpaper}, the LOFAR software suite\footnote{\url{https://www.astron.nl/lofarwiki/doku.php?id=public:user_software:start}} and PIMA\footnote{\url{http://astrogeo.org/pima/pima_user_guide.html}} \citep{pima}. The packages were compared on their abilities at that time to serve as a fully capable VLBI data processing package, with focus on the necessary steps to process correlated VLBI data into scientific results. 
To compare the packages a list of five aspects was made: reliability, flexibility, future prospects, user access and pipeline readiness. Each aspects included several requirements, which were weighted highest for critical requirements, to lowest for optional requirements. For each requirement the packages received scores between 1 (lowest) and 5 (highest). With this information a weighted average was calculated for each aspect. This resulted in an overall score with standard deviation for each package, see Table~\ref{table:software_comparison}. The standard deviation is indicative of how much spread there is in the grades, and therefore the need for significant adjustments to meet some of the requirements. A lower standard deviation implies that overall there is less work to be done to meet individual requirements, while a high standard deviation means that though some requirements are met, significant discrepancies exist for others, which implies more work.
The highest score was for CASA ($4.5\pm0.8$), with AIPS a good runner-up ($3.9\pm1.2$). The other packages scored well below 3, only the LOFAR toolkit scored 3.2, but required significant adjustments given the very large difference in frequency, and it was not a singular software package but rather a collection of tools and packages. In further assessment with the development teams at NRAO it became clear that CASA would take priority for their future work. Combined with the better score, this led to the choice to use CASA as the basis for further development.

\setcounter{table}{0}
\renewcommand{\thetable}{A\arabic{table}}

\begin{table}[!h]
    \centering
    \begin{tabular}{lcc}
    \hline
    \hline
    Package & Mean & Standard Dev. \\
    \hline
    AIPS    & 3.9   & 1.2 \\
    Miriad  & 2.3   & 1.0 \\ 
    HOPS    & 2.3   & 1.1 \\
    CASA    & 4.5   & 0.8   \\
    LOFAR   & 3.2   & 1.3   \\
    PIMA    & 2.3   & 1.6   \\
    \hline
    \end{tabular}
    \caption{Weighted mean scores and standard deviation for all packages assessed for the software comparison. The scoring was done on a range of requirements, with 0 being the lowest and 5 the highest score. Weighting was applied based on the necessity of each requirement.}
    \label{table:software_comparison}
\end{table}

\end{document}